\documentclass{elsart}

\usepackage{graphicx}

\usepackage{amssymb}
\journal{Physics Letters A}

\begin{document}
\begin{frontmatter}

\title{Probing internal bath dynamics by a Rabi oscillator-based detector}
\author{Murat \c{C}etinba\c{s}\corauthref{cor}},
\corauth[cor]{Corresponding author.} 
\ead{cetinbas@sfu.ca}
\author{Joshua Wilkie}
\ead{wilkie@sfu.ca}
\address{Department of Chemistry, Simon Fraser University, Burnaby, British Columbia V5A 1S6, Canada}

\begin{abstract}
By exact numerical and master equation approaches, we show that a central spin-1/2 can be configured to probe internal bath dynamics. System-bath interactions cause Rabi oscillations in the detector and periodic behavior of fidelity. This period is highly sensitive to the strength of the bath self-interactions, and can be used to calculate the intra-bath coupling.
\end{abstract}

\end{frontmatter}


\section{Introduction}
\label{intro}
Recent identification of potential quantum computer (QC) architectures \cite{QC} with long external decoherence times
has shifted attention toward the negative effects of static internal structural flaws \cite{GS}. Isolated flawed QCs 
and their immediate local environments can in many cases be mapped onto self-interacting spin-bath models, and as a consequence
spin baths are increasingly a focus of attention \cite{GS,Tess,Mil,CW}. Studies have 
reported a number of interesting dynamical effects such as suppression of decoherence
with increasing bath self-interaction \cite{Tess,Mil,CW,Sanz}, and large coherent shifting of subsystem dynamics \cite{CW}. 

Here we show that the coherent shift can be used to configure a single qubit detector to measure the strength of bath self-interactions.
Structural flaws in a QC arise due to variance in one-body parameters or as a consequence of residual two-body interactions,
or even as a result of strong interactions with other local impurities. One body flaws can be readily detected and corrected in
many cases. Two-body interactions are more difficult to deal with and their effects are difficult to predict. Knowledge of the
strength of the two-body interactions is obviously a prerequisite for eliminating their effects. It has been observed that the
coherent shift is sensitive to the strength of bath self-interaction \cite{CW}. We show that a single qubit can be configured
to show large amplitude oscillations in fidelity when a shift is present, and that the period of the oscillations is highly
sensitive to the strength of the bath self-interaction. From the period of the oscillation one can extract a quantitative 
estimate of the self-interaction strength. While we are not aware of any existing error correction schemes designed specifically for two-body flaws one should be simple to devise. Moreover, it should be possible to apply the basic ideas behind our detector set-up in more general contexts such as optical impurities in solids where knowledge of bath self-interaction could be important.

An engaging aspect of spin-bath models is the ease with which exact results can be obtained for the low temperature regime 
relevant to quantum computing. Unfortunately, exact simulations are necessarily limited to small baths. In our study this 
limitation is not crucial but in more general studies it might be
very important. Consequently, we ideally would like to have a second independent way of obtaining open system dynamics.
While uncoupled spin baths and uncoupled oscillator baths can display similar dynamics\cite{Pech}, standard boson-bath models\cite{Gard,Harmo,NMQSD,Lamb} are
inapplicable here. This is due to the fact that the bath self-interactions are non-negligible in our model. We therefore choose to also consider simulations obtained using a recently developed approximate master equation (ME) \cite{SRA,SRA2,SRA3}, which takes into account bath self-interactions via a mean-field approach.

Our QC model is based on a Josephson junction (JJ) based proposal \cite{Nori} which has an extremely long external decoherence time\cite{Nori,Makhlin}. 
Recent experiments report \cite{exp-JJ,exp-JJ2} that the primary cause of decoherence of such JJ qubits is static interaction with
some unidentified local two-level systems (TLSs) \cite{TLS,Rabi-mech}. In one case \cite{exp-JJ2} externally induced Rabi oscillations
in a qubit were significantly distorted at times much shorter than the external decoherence time and the existence of the TLSs was
posited as the cause. Our recent results \cite{CW} suggest that the distortion might be caused by a coherent shift which could
arise just from flaws and residual interactions amoung the qubits. In any case the
possible TLS can be viewed as distorted auxilliary qubits \cite{TLS}. The computer and TLSs can be mapped onto a spin-bath and this
is the model we address.

We let one detector qubit interact with a statically flawed QC, and our goal is to show that the strength of the 
bath self-interaction can be measured. Our system is thus a single two level system and our bath consists of flawed qubits and other interacting
TLSs. We prepare our detector in a superposition state which will merely undergo phase evolution in the absence of coherent
shifting. When shifting is present the populations and fidelity oscillate. The strength of the bath self-interaction is 
extracted from the period of these oscillations. We use both exact and approximate master equation simulation
methods.

While the phenomenology of decoherence and dissipation dominate the literature of open quantum systems\cite{Tess,Mil,CW,Sanz,Pech,Gard,Harmo,NMQSD,Lamb,SRA,SRA2,SRA3} the possible existence of coherent shifting has been acknowledged but not widely explored. Combinations of observables like fidelity 
and purity can be used to distinguish these coherent sorts of errors from non-unitary effects like decoherence and dissipation \cite{CW}. The effect of
such a shift can be non-trivial and it can qualitatively alter the system evolution \cite{CW}. In our detector, for example, we will see that pure phase evolution is changed to large amplitude Rabi oscillations. 

Distinguishing the unitary part of open system-dynamics from non-unitary parts is of particular importance, because unitary errors induced by the bath should be easily correctable. The exact Nakajima-Zwanzig ME \cite{Zwan} contains terms which result in both
Markovian and non-Markovian distortions. The Markovian term consists of Hermitian corrections to the subsystem Hamiltonian and
clearly causes unitary errors.
Unfortunately, this Markovian correction cannot always be identified with the physical coherent shift. Different choices for the projection operator \cite{nonher} lead to different Markovian corrections.
It is possible in principle that a specific projection operator will have the property that the Markovian corrections are
precisely equal to the physical shift \cite{SRA,SRA2,SRA3}. The identity of this projection operator is a matter of some interest since it is
in a sense the most physical. Furthermore, the form of the Nakajima-Zwanzig master equation \cite{Zwan} which emerges from
this choice of the projection operator would be a favorable starting point for subsequent approximations.

A recently developed approximate ME \cite{SRA,SRA2,SRA3} relies on a projection operator for which the unitary corrections
consist of a shift $\hat{H}_S\rightarrow \hat{H}_S+\bar{B}\hat{S}$, where 
$\bar{B}$ denotes the canonical average of the bath coupling operator $\hat{B}$ and $\hat{S}$ is the system coupling operator. 
The non-Markovian terms are all non-unitary and so agreement of this ME with exact calculations would
strongly support this choice of the projection operator. We show that the predictions of the ME are indeed quite good and
hence that this the projection operator is the most physical. The accuracy of the ME in this instance is of some 
interest since it also preserves positivity and has no adjustible parameters.

Organization of this Letter is as follows. In Section 2, we explain the mathematical details of our model. In Section 3
we explain our exact simulation method. Section 4 reviews the derivation of the exact Nakajima-Zwanzig ME
and the approximations which lead to a solvable positivity-preserving non-Markovian ME with no adjustable parameters.  A numerical strategy for solving the master equation is also reviewed.
In Section 5, we present and discuss our exact numerical and master equation results. 

\section{Self-Interacting Spin-Bath Model}
\label{exact}
Our self-interacting spin-bath model, representing a flawed QC core consists of $N+1$ two level systems, is given by the following total Hamiltonian
\begin{equation}\label{Ham}
\hat{H}=\hat{H}_{S}+\hat{S}\hat{B}+{\hat H}_{B}
\end{equation}
where the first term represents the system Hamiltonian
\begin{equation}
\hat{H}_{S}=-\frac{1}{2}B_{0}^{z} \hat{\sigma}_{z}^{(0)}
\end{equation}
the second term is a system-bath interaction operator in system $\hat{S}$ and bath $\hat{B}$ spaces
\begin{equation}
\hat{S}\hat{B}=\hat{\sigma}_{x}^{(0)} \sum_{i=1}^{N}\lambda_{i} \hat{\sigma}_{x}^{(i)}
\end{equation}
and the bath Hamiltonian is given by 
\begin{eqnarray}
{\hat H}_{B}&=&-\frac{1}{2}\sum_{i=1}^{N} \left( B_{i}^{x}\hat{\sigma}_{x}^{(i)} 
+ B_{i}^{z} \hat{\sigma}_{z}^{(i)} \right) \nonumber \\
&& ~~~~~~~~~~~~+ \sum_{i=1}^{N-1}\sum_{j=i+1}^{N} J_{x}^{i,j} \hat{\sigma}_{x}^{(i)}\hat{\sigma}_{x}^{(j)}
\end{eqnarray}
Note that $\hat{\sigma}_{\alpha}^{i}$ are standard Pauli spin-1/2 operators with $\alpha=\{x,y,z\}$ and the index $i$ labels the qubits. 

Hamiltonian (\ref{Ham}) is a variant of an isolated flawed QC model recently introduced to investigate the effects of one-qubit static imperfections and two-qubit residual interactions on the performance of QCs\cite{GS}. The actual internal dynamical effects have been explored \cite{CW}
and while decoherence and dissipation are present, the principle errors were caused by a coherent shift. 

Here qubit 0 will be employed as our detector while the rest represent some combination of flawed qubits and TLSs. Flaws will be modeled by 
choosing the parameters of $\hat{H}_B$ randomly from some preset distribution. Note also that the bath qubits and TLSs interact among themselves.

\subsection{Parameters}
The parameters used in our numerical study are based on a Josephson charge-qubit QC\cite{Nori,Makhlin} proposal for which the experimentally accessible single qubit energy to perform a one-qubit rotation is $B_{0}^{z}=1 \epsilon$ with $\epsilon=200$ mK. Since all qubits in Hamiltonian (\ref{Ham}) can be viewed as components of the same QC, the fluctuations in single qubit parameters should only differ from $B_{0}^{z}$ by a detuning parameter, which we set to the value $\delta=0.4\ \epsilon$. Thus, we add noise
to all qubits except for our detector by choosing $B_i^z \in [ B_0^z-\delta/2,B_0^z+\delta/2 ]$. We also make an allowance for $x$-type errors by choosing
$B_i^x\in [B_0^z-\delta/2,B_0^z+\delta/2 ]$. Following Ref. \cite{GS} residual two-body interactions are included by randomly choosing $\lambda_{i}$ and $J_{x}^{i,j}$ uniformly from  $\lambda_{i} \in \ [-\lambda,\: \lambda]$ and  $J_{x}^{i,j} \in \ [-J_{x},\: J_{x}]$, respectively. While we considered a number of intra-bath coupling strengths $J_{x}=0.00, \: 0.15, \: 0.50, \: 1.00, \:2.00$ in units of $\epsilon$, to explore the integrable to chaotic transition, we considered only one system-bath coupling value, which corresponds to the experimental value for a two-qubit rotation, $\lambda=0.05\ \epsilon$. In all our calculations temperature is kept constant at $kT=0.25\ \epsilon$ and time step of integration was $\Delta t=0.2~\hbar/\epsilon$ s.

Thus, the parameters are set so that all qubits are part of the same computer, but one qubit does not possess any one-body flaws and we use this as our detector. Note also that in a multi-qubit JJ QC, the existence of the aforementioned TLSs is expected. In this study, we will not attempt to explicitly distinguish between qubits and TLSs because the nature of the TLSs is currently unknown and it is not possible to derive TLS parameters from a
first principle. Nevertheless, the same form of pseudospin  Hamiltonian used for the qubits is also assumed for TLSs \cite{Nori2} and they both have comparable decoherence times \cite{Nori2}. 

\subsection{Initial Conditions}
We assume that the initial state of the composite system is of product form
\begin{equation}
\hat{\rho}(0)=\hat{\rho}_{S}(0)\otimes \hat{\rho}_{B}(0)
\end{equation}
\noindent
where $\hat{\rho}_{S}(0)=|\psi(0) \rangle \langle \psi(0)|$ is the initial pure reduced density of the system, and $\hat{\rho}_{B}(0)$ is the canonical bath density. We chose the
system initial state as $|\psi(0)\rangle=(|0\rangle+|1\rangle)/\sqrt{2}$. In the absence of interactions with the bath this state of the detector will
undergo phase evolution only. But Rabi oscillations will appear when there are interactions with the bath. 

\section{Exact Numerical Approach}
Exact numerical solutions are obtained by exploiting the low temperature regime for the bath degrees of freedom. Thus we approximate the initial bath density as
\begin{equation}
\hat{\rho}_{B}(0) =\sum_{n=1}^{n_{cut}} p_{n} |\phi_{n}^{B} \rangle  \langle \phi_{n}^{B} |\label{sum}
\end{equation}
where $p_{n}=\exp{(-E_{n}/kT)}/\sum_{m=1}^{n_{cut}}\exp{(-E_{m}/kT)}$ are the thermal populations, and $\hat{H}_{B} | \phi_{n}^{B} \rangle = E_{n} | \phi_{n}^{B} \rangle$. Note that the sum is over only the thermally populated, lowest energy, eigenstates of the bath. Hence, $n_{cut}$ is a cutoff such that states $n_{cut}+1$ and higher are unoccupied for our given fixed temperature. Even though the density of bath states and thus the number of populated states varies slowly with $J_x$ for a fixed temperature,  $n_{cut}=20$ was sufficient for all cases in our calculations. 
 
We calculate the exact reduced density $\hat{\rho}_{S}(t)$ at time $t$ via
\begin{equation}
\hat{\rho}_{S}(t)=\sum_{n=1}^{n_{cut}} p_{n}~{\rm Tr}_{B} [ 
|\Psi_{n}(t) \rangle \langle \Psi_{n}(t)|],
\label{TDen2}
\end{equation}
\noindent
Here the $|\Psi_{n}(t) \rangle$ evolve according to the Schr\"{o}dinger equation
\begin{equation}
i\hbar \frac{d}{dt}|\Psi_n(t)\rangle= \hat{H}|\Psi_n(t)\rangle,\label{SE}
\end{equation}
\noindent
and all initial states are of the form 
$|\Psi_n(0)\rangle=|\psi(0)\rangle\otimes |\phi_{n}^{B} \rangle$. We used a Lanczos algorithm\cite{ARPACK} for exact 
diagonalization of the bath Hamiltonian, for $N=10$ qubits, and an eighth order variable stepsize Runge-Kutta method\cite{RK} for the numerical integrations of Eq. (\ref{SE}). 

From the exact reduced density various measures (discussed in Section 5) can be calculated to determine the extent of decoherence, dissipation, and shifting.

\section{Master Equation Approach}
\label{Master}
Here we begin by reviewing the basic ideas behind the Nakajima-Zwanzig projection operator technique and show how it leads to an exact but unsolveable master equation. Next we explain how a mean field approximation can be used to
obtain an approximate but solveable master equation with no adjustible parameters.

\subsection{Exact Nakajima-Zwanzig Master Equation}
Exact quantum dynamics in density operator space obeys the Liouville-von Neumann equation
\begin{equation}
i\frac{d}{dt}\hat{\rho}(t)=\hat{L}\hat{\rho}(t)
\label{LW}
\end{equation} \noindent
where the Liouville operator is $\hat{L}=(1/\hbar)[\hat{H}_{S}+\hat{S}\hat{B}+\hat{H}_{B}, \cdot~]$. The Nakajima-Zwanzig approach begins by defining a projection operator
$\hat{P}$ such that $\hat{P}\hat{\rho}(t)$ is proportional to the reduced density $\hat{\rho}_S(t)$. Once a $\hat{P}$ is defined, then its complement is
$\hat{Q}=\hat{1}-\hat{P}$. Operating on both sides of the Liouville-von Neumann equation with $\hat{P}$ and $\hat{Q}$ yields
\begin{eqnarray}
i\frac{d}{dt} [\hat{P}\hat{\rho}(t)]&=&\hat{P}\hat{L}\hat{P} ~\hat{P}\hat{\rho}(t)+\hat{P}\hat{L}\hat{Q}~\hat{Q}\hat{\rho}(t)
\label{s1} \\
i\frac{d}{dt} [\hat{Q}\hat{\rho}(t)]&=&\hat{Q}\hat{L}\hat{P}~\hat{P}\hat{\rho}(t)+\hat{Q}\hat{L}\hat{Q} ~\hat{Q}\hat{\rho}(t)
\label{s2}.
\end{eqnarray}
\noindent
This coupled set of equations can then be solved to find an equation for $\hat{P}\hat{\rho}(t)$:
\begin{eqnarray}\label{prho}
i\frac{d}{dt} \hat{P}\hat{\rho}(t)&=&\hat{P}\hat{L}\hat{P}\ \hat{P}\hat{\rho}(t)+\hat{P}\hat{L}\hat{Q}\ e^{-i\hat{Q}\hat{L}\hat{Q}t}\ \hat{Q}\hat{\rho}(0) \nonumber \\ 
&-&i\int_{0}^{t}dt^{\prime}\hat{P}\hat{L}\hat{Q}\ e^{i\hat{Q}\hat{L}\hat{Q}(t^{\prime}-t)}\ \hat{Q}\hat{L}\hat{P}\ \hat{P}\hat{\rho}(t^{\prime}) \label{GME}
\end{eqnarray}
Tracing over the bath degrees of freedom at this stage would lead to a general master equation for the reduced 
density. 

The properties of the resulting master equation will in part be determined by the choice of $\hat{P}$. In particular, it may be 
possible to choose a $\hat{P}$ such that the non-Markovian term generates non-unitary evolutions only. 
Assuming that the system and thermal bath are initially uncorrelated (i.e. $\hat{\rho}(0)=\hat{\rho}_{S}(0)\otimes \hat{\rho}_{B}(0)$) and that the number of 
bath degrees of freedom is large enough so that the system does not significantly alter the equilibrium state of the bath, then the
total density $\hat{\rho}(t)$ at time $t$ will be approximately given by $\hat{\rho}_S(t)\otimes \hat{\rho}_B(0)$. Thus, it seems to make sense to
define the projection operator $\hat{P}$ via
\begin{equation}
\hat{P}\hat{\rho}(t)=\hat{\rho}_{S}(t)\otimes \hat{\rho}_{B}(0).\label{Ptrue}
\end{equation}
It should be noted that this was not the actual projection operator employed by Zwanzig\cite{Zwan}. This particular definition has
however been widely employed in the literature, and it is well known to be of non-Hermitian type\cite{SRA,SRA2,SRA3,nonher}. With
this $\hat{P}$ the master equation reduces to the form
\begin{eqnarray}
\frac{d}{dt}\hat{\rho}_S(t)&=&-\frac{i}{\hbar}[\hat{H}_S+\bar{B}\hat{S},\hat{\rho}_{S}(t)] \\
&-&\int_0^{t}dt^{\prime}~{\rm Tr}_B\{ \hat{\rho}_B(0) \hat{L}\hat{Q}e^{i\hat{Q}\hat{L}\hat{Q}(t^{\prime}-t)}\hat{Q}\hat{L} \}\hat{\rho}_S(t^{\prime}) \nonumber
\end{eqnarray}
in which we see the emergence of the coherent shift $\hat{H}_S \rightarrow \hat{H}_S+\bar{B}\hat{S}$. The problem with this master equation is that 
evaluation of the memory kernel is extremely difficult.

\subsection{Non-Markovian Mean Field Master Equation}

More useful, although approximate, equations are obtained by making approximations for the memory kernel. Here we explore a mean
field approach. 

Since $\hat{P}$ and $\hat{Q}$ are not Hermitian the operator $\hat{Q}\hat{L}\hat{Q}$, mediating correlations between system and bath, is not Hermitian either \cite{SRA2,SRA3}. Assuming that $\hat{Q}\hat{L}\hat{Q}$ has a complete complex spectral decomposition, then for evolution forward in time $t \ge 0$ we write
\begin{equation}
\hat{Q}\hat{L}\hat{Q} = \sum_{\mu}(\omega_{\mu}-i\gamma_{\mu})|\phi_{\mu}^{+})(\Phi_{\mu}^{+}|
\label{qlq}
\end{equation}
where $|\phi_{\mu}^{+})$ and $(\Phi_{\mu}^{+}|$ are the right and left eigenvectors, respectively. Substituting (\ref{qlq}) back into (\ref{GME}) and tracing over the bath gives
\begin{eqnarray} 
\frac{d}{dt}\hat{\rho}_{S}(t)=-\frac{i}{\hbar} \left[ \hat{H}_{S}+\bar{B}\hat{S}, \hat{\rho}_{S}(t)\right]
-\int_{0}^{t} dt^{\prime}\hat{{\cal K}}(t-t^{\prime})\hat{\rho}_S(t^{\prime})
\label{exact2}
\end{eqnarray}
\noindent
where
\begin{eqnarray}
\hat{{\cal K }} (t-t^{\prime})= &&\sum_{\mu}e^{-i(\omega_{\mu}-i\gamma_{\mu})(t-t^{\prime})} \nonumber \\  & \times & {\rm Tr}_{B}\{ \hat{\rho}_B(0)\hat{L}\hat{Q} |\phi_{\mu}^{+})(\Phi_{\mu}^{+}|\hat{Q}\hat{L} \}.
\label{memop2}
\end{eqnarray}

Further progress now requires some form of approximation for above sum since the eigenspectrum cannot be obtained in practice.
A mean field approach seems logical if $|\phi_{\mu}^{+})$ and $(\Phi_{\mu}^{+}|$ are random functions as they would be for a chaotic 
bath. Further assuming that the eigenvalues $\omega_{\mu}-i\gamma_{\mu}$ are statistically independent of the eigenvectors
then suggests that the sum should be replaced by its average
\begin{displaymath}
\langle e^{-i(\omega_{\mu}-i\gamma_{\mu})t} \rangle \sum_{\mu} |\phi_{\mu}^{+})(\Phi_{\mu}^{+}|=\langle \cos({\omega t})e^{-\gamma t}\rangle \hat{1}_S\otimes \hat{1}_B.
\end{displaymath}
Defining $W(t)=\langle \cos({\omega t})e^{-\gamma t}\rangle$ then we obtain
\begin{equation}
\frac{d}{dt}\hat{\rho}_{S}(t)=-\frac{i}{\hbar} [\hat{H}_{S}+\hat{S}\bar{B}, \hat{\rho}_{S}(t)]  \label{sra}  
-\int_{0}^{t}dt^{\prime}W(t-t^{\prime}) 
\hat{\cal L}_{D}\rho_{S}( t' ).
\end{equation}
Here $\hat{\cal L}_{D}=(C/\hbar^{2})\{ [\hat{\rho}_{S}\hat{S},\hat{S}]+[\hat{S},\hat{S}\hat{\rho}_{S})]  \}$ is a dissipative Lindblad-Kossakowski superoperator \cite{dsg}, and
and $C={\overline{B^2}}- {\bar{B}}^{2}$ is the variance of the bath coupling operator. Here the non-Markovian term is purely dissipative
and the only unitary effects arise from the shift $\bar{B}\hat{S}$. 

In the Markovian limit (i.e. when memory effects are negligible)
\begin{eqnarray} \nonumber
\int_{0}^{t} dt' W(t-t')\hat{\cal L}_{D}\hat{\rho}_{S}( t' ) &\simeq&
\int_{0}^{\infty} dt'  W( t' )\hat{\cal L}_{D}\hat{\rho}_{S}( t ) \\ 
&\sim& \tau_{B} \hat{\cal L}_{D}\hat{\rho}_{S}( t ) \label{Markov}
\end{eqnarray}
\noindent
where $\tau_{B}$ is the (short) memory relaxation time. In this limit Eq. (\ref{sra}) reduces to the standard completely-positive-dynamical-semigroup (CPDS) form\cite{dsg}
\begin{equation}
\frac{d}{dt}\hat{\rho}_{S}(t)=-\frac{i}{\hbar}[\hat{H}_{S}+\hat{S}\bar{B}, \hat{\rho}_{S}(t)]  \label{sra-Mar}  
-\tau_{B}\hat{\cal L}_{D}\hat{\rho}_{S}(t).
\end{equation}

We however will employ the non-Markovian form since memory effects are important in our system.
Analysis of the spectral properties of $\hat{Q}\hat{L}\hat{Q}$ suggests that an appropriate memory function for chaotic baths is of the form\cite{SRA3}
\begin{eqnarray}
W(t)=[1-\frac{4}{3\pi} (pt)^1+\frac{1}{8}(pt)^2-\frac{4}{45\pi}(pt)^3 ] e^{-(q t)^2/8}
\label{W2}
\end{eqnarray}
where expressions for the quantities $p$ and $q$ are presented elsewhere\cite{CW2}. Here $W(t)$ is positive and real.

It is important to note that the master equation (\ref{sra}) preserves positivity \cite{SRA,SRA2,SRA3} for our positive 
memory function, and under certain conditions outlined in \cite{Budini} it also preserves complete positivity. The master equation is free of adjustable parameters and thus it can be solved numerically without any further approximation. Hence the master equation (\ref{sra}) can be viewed as a non-Markovian generalization of CPDS theory \cite{dsg}, which draws a sharp distinction between unitary and non-unitary contributions to the system dynamics. New types of projection operators, leading to other non-Markovian generalizations of CPDS theory \cite{dsg} are discussed in \cite{Breuer}. Agreement of our master equation with exact calculations would then suggest that Eq. (\ref{Ptrue}) is indeed the physically meaningful choice for $\hat{P}$. 

\subsection{Numerical Solution of Master Equation}
We use a recently developed numerical method for solution of non-Markovian equations\cite{TU}. The method is implemented by converting integro-differential equations to ordinary differential equations. First, define a space-like time variable $u$ and a smoothed density function
\begin{equation}\label{SMO}
\hat{\chi}(t,u)=f(u)\int_0^tdt'~W(t-t'+u)\hat{\rho}(t'),
\end{equation}
where $f(u)$ is a damping function with $f(0)=1$. Then, substituting Eq. (\ref{SMO}) into Eq. (\ref{sra}) we obtain two coupled ordinary differential equations for $\hat{\rho}(t)$ and $\hat{\chi}(t,u)$
\begin{eqnarray}
\frac{d}{dt}\hat{\rho}_{S}(t)&=&-\frac{i}{\hbar}[\hat{H}_{S}+\hat{S}\bar{B},\hat{\rho}_{S}(t)] \nonumber \\
&-&\frac{C}{\hbar^{2}}\{[\hat{\chi}(t,0)\hat{S},\hat{S}]+[\hat{S},\hat{S}\hat{\chi}(t,0)]\},  \label{tu1} \\
\frac{d}{dt}\hat{\chi}(t,u)&=&f(u)W(u)\hat{\rho}_{S}(t) \nonumber \\
&+&\frac{\partial \hat{\chi}(t,u)}{\partial u}-\frac{f'(u)}{f(u)}~\hat{\chi}(t,u)\label{tu2}
\end{eqnarray}
which can be solved using standard methods.

For the spin--spin-bath model of our study these equations have the form ($\hbar=1$)
\begin{eqnarray}
\frac{d}{dt}\hat{\rho}_{S}(t)&=&-i[-\frac{B_0}{2}\hat{\sigma}_z^{(0)}+\bar{B}\hat{\sigma}_x^{(0)},\hat{\rho}_{S}(t)] \nonumber \\
&-&2C\{ \hat{\chi}(t,0)-\hat{\sigma}_{x}^{(0)}\hat{\chi}(t,0)\hat{\sigma}_{x}^{(0)} \} \label{stu1}\\
\frac{d}{dt}\hat{\chi}(t,u)&=&e^{-g u^2}W(u)\hat{\rho}(t) \nonumber \\ 
&+&\frac{\partial \hat{\chi}(t,u)}{\partial u}+2g u~\hat{\chi}(t,u)\label{stu2},
\end{eqnarray}
where $\bar{B}=\bar{\Sigma}_x$ and $C_{x}=\overline{\Sigma^{2}_{x}} -\bar{\Sigma}_{x}^{2}$. Here ${\Sigma}_{x}=\sum_{k=1}^{N}\lambda_{k}\hat{\sigma}_x^{(k)}$ and the overbar denotes a canonical average over the initial bath state. The parameters of the memory function, $C_{x}$, and $\bar{\Sigma}_{x}$ are calculated using exact energies and eigenvectors of the bath Hamiltonian. 

We found that a damping function of the form $f(u)=e^{-g u^2}$ with $g=  9.9/[(n-l)\Delta t]^2$, as suggested in Ref. \cite{TU}, was accurate for our problem. Using $W(u)=W(|u|)$ for negative values of $u$ the auxiliary differential equation (\ref{stu2}) is solved on a grid of points $u_j=(n+l-j)\Delta t$ with $j=1,\dots, n$ and $l=int(.338n)$. Here $\Delta t=0.2~\hbar/\epsilon$ is the time-step employed in the dynamics. Partial derivatives with respect to $u$ are calculated by using a discrete-variable\cite{DVR} matrix representation. Converged results were obtained for $n=40$ grid points. The ordinary differential equations (\ref{stu1}) and (\ref{stu2}) were integrated by an eighth order variable stepsize Runge-Kutta routine\cite{RK}. 

\section{Results}
We use purity and fidelity to detect and quantify deviations from pure phase evolution in our detector. Purity, defined by ${\cal{P}}(t)={\rm Tr}_{S}[ \hat{\rho}_{S}(t) ]^{2}$, is a good measure of decoherence and dissipation since it is insensitive to the coherent shift.  
For pure initial states, such as we will employ, the dynamics of the system will ideally have a purity equal to 1. This would be the case if interaction 
with the bath causes only coherent shifting but no non-unitary effects.

Fidelity, defined by ${\cal{F}}(t)={\rm Tr}_{S}[ \hat{\rho}_{S}(t)\hat{\rho}_{S}^{ideal}(t) ]$, is sensitive to both unitary and 
non-unitary effects. Here $\hat{\rho}_{S}^{ideal}(t)$ is the system density at time $t$ in the absence of system-bath interactions. Fidelity measures the proximity of an open system's evolution to its free evolution. Hence, for pure initial states the ideal value of fidelity is also 1. Comparison of the magnitudes of purity
and fidelity gives an indictation of the presence of coherent shifts.

Recall now that we have chosen a initial state $(|0\rangle +|1\rangle)/\sqrt{2}$ for the system. In the absence
of coherent shifting this state will undergo only phase evolution without significant population transfer. In the presence of coherent shifting we 
expect that the system will undergo large amplitude Rabi oscillations. We thus also look at the populations of the states to additionally
monitor the emergence of the coherent shift process.

\begin{figure}
\centering
\includegraphics*[scale=0.3]{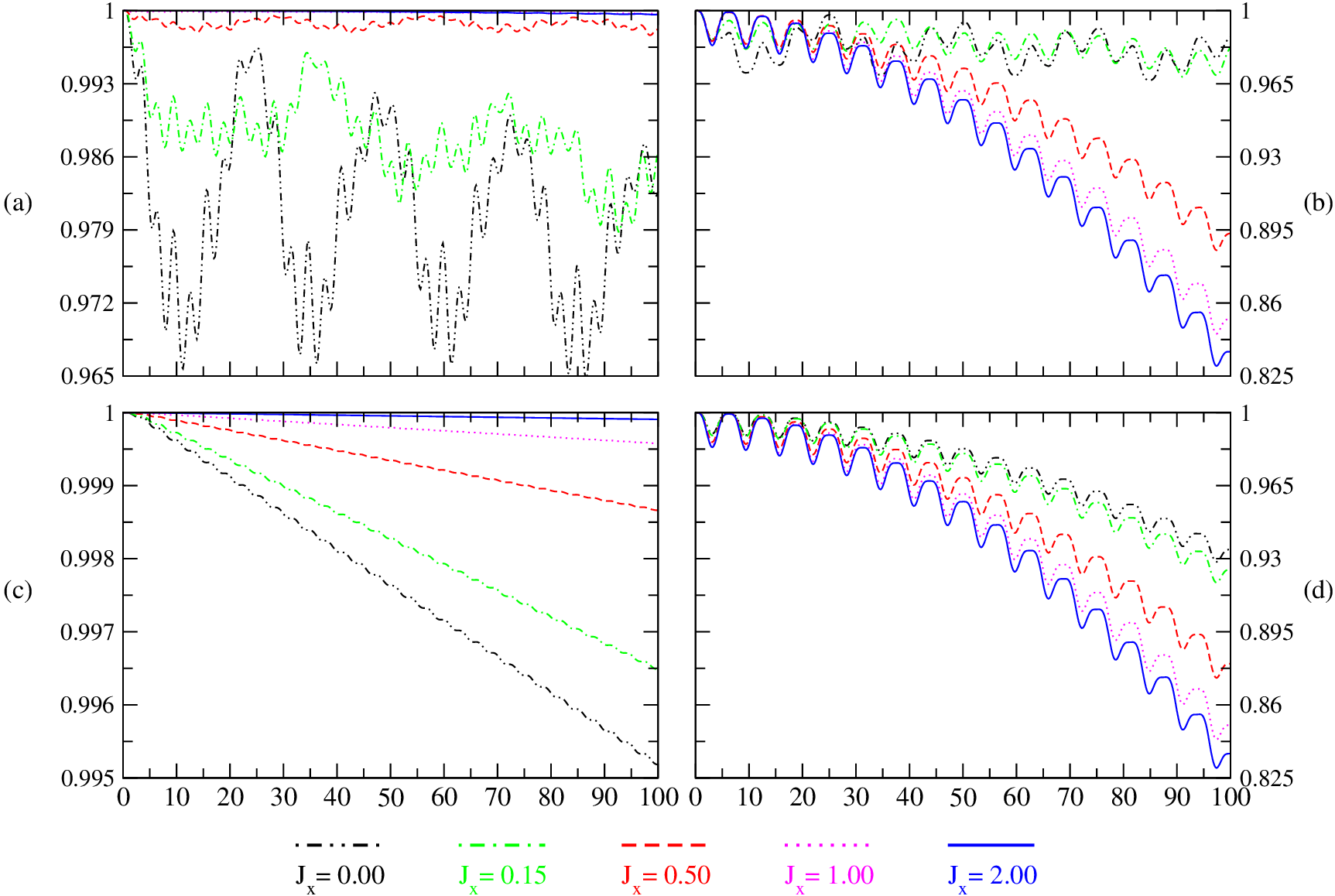} 
\caption{(Colour online) Exact numerical results are plotted in (a) for ${\mathcal{P}}(t)$ and in (b) for ${\mathcal{F}}(t)$. Master equation results are plotted in (c) for ${\mathcal{P}}(t)$ and in (d) for ${\mathcal{F}}(t)$. Five different intra-bath couplings $J_{x}=0.00,0.15,0.50,1.00,2.00$ are shown in each subplot.} 
\label{Fig1}
\end{figure}

\subsection{Short Time Dynamics}
The short time behaviors of ${\mathcal{P}}(t)$ and ${\mathcal{F}}(t)$ are seen in Figure 1 for both exact and master equation results for five different intra-bath couplings, $J_{x}$. The systematic improvement in ${\mathcal{P}}(t)$ with increasing $J_{x}$ is observed for both exact and master equation calculations. For the integrable bath i.e. $J_{x}=0.00$ the decoherence is at its maximum. Above $J_{x}=0.15$ chaos sets in and the decoherence is systematically reduced. For strong chaos with $J_{x}=2.00$ the decoherence is almost totally suppressed. This result is in agreement with earlier studies\cite{Tess,Mil,CW,Sanz} in which the bath chaos is predicted to reduce decoherence and dissipation. 

Partial recurrences of the purity are observed in the exact calculation in the integrable regime. This is obviously a consequence of the 
small bath size and the effect is not observed in the master equation results. Agreement between the master equation and exact calculation
clearly improves with increasing $J_x$, and this is a consequence of the fact that the dynamics becomes more Markovian with greater
bath chaos. Markovian character requires a separation of relaxation times for the system and bath degrees of freedom \cite{Silbey}. A chaotic
bath can relax internally, but the only available relaxation mechanism for an integrable bath is through the system. Thus, chaotic baths
tend to be more Markovian. Since we assumed the bath was chaotic in its derivation we should not be surprised that the ME is more 
accurate for larger $J_x$.

The dissipation term in Eq. (\ref{sra}) is preceded by a factor which is equal to the canonical variance in the bath
coupling operator. This variance is a measure of the number of bath states which can contribute to decoherence and dissipation of the
system. Fig. 2(b) shows that this variance declines with $J_x$ and so this in part explains why Fig. 1 shows an improvement in 
purity with increasing $J_x$. It has been argued elsewhere\cite{Tess} that the off-diagonal matrix elements of the bath
coupling operator tend to vanish as the chaotic regime is approached. This lack of selection rules is largely responsible
for the decline in the variance.

\begin{figure}
\centering
\includegraphics*[scale=0.3]{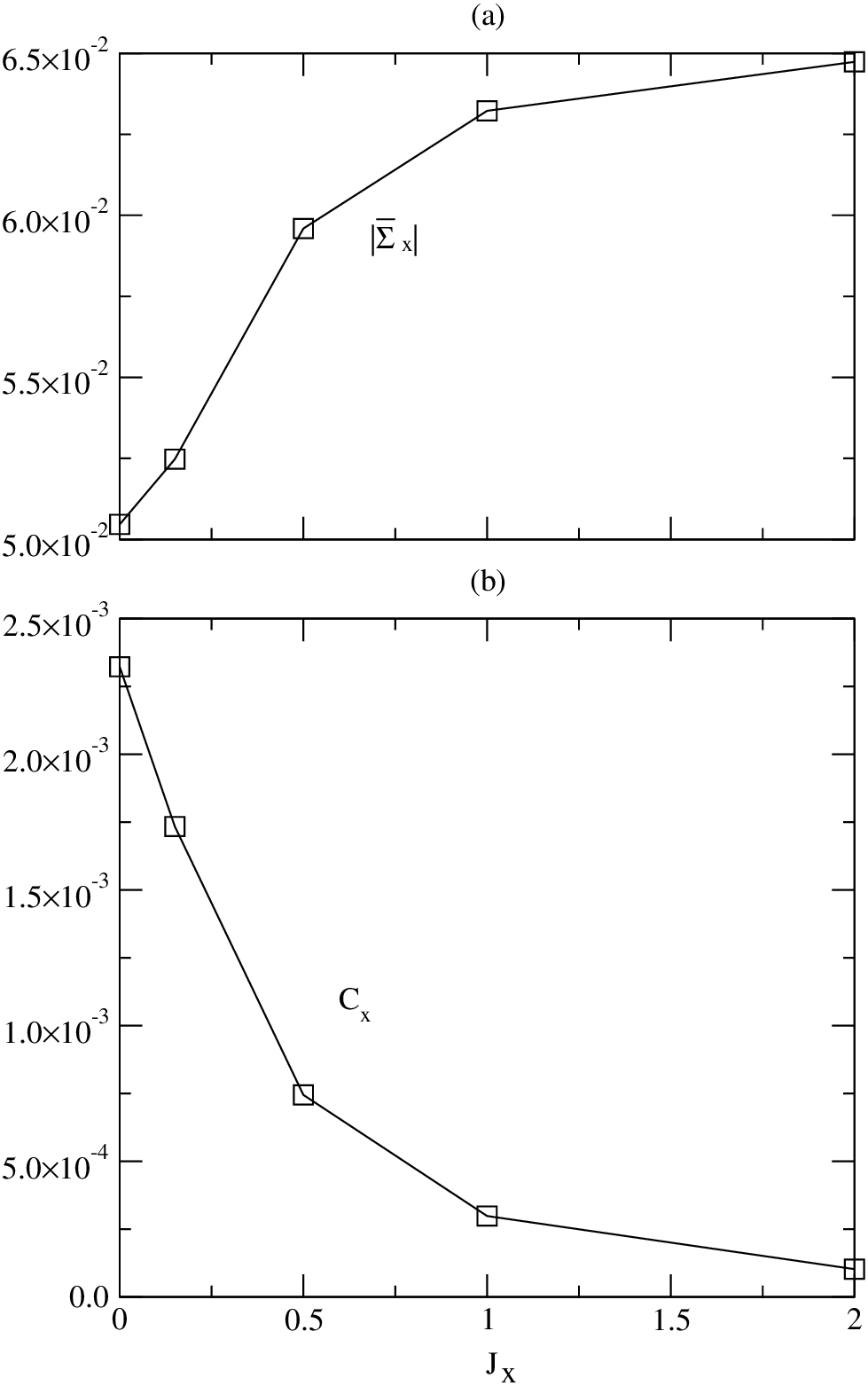} 
\caption{Canonical average (a) and variance (b) of bath coupling operator plotted vs. $J_x.$}
\label{Fig2}
\end{figure}

The fidelity plots of Fig. 1(b) show the presence of much larger deviations from unity than that seen in the purity plots. This supports the 
conclusion that large coherent shifts are again occuring as they did in our previous study\cite{CW}. This shift is clearly much more 
harmful than decoherence or dissipation as an potential error source for quantum computation \cite{CW}. We will see that the coherent shifting can be put to a good purpose, however. It is also worth noting
that while the coherent shift vanishes in some standard models of decoherence, it is nonzero in general for self-interacting baths. Thus, we expect that
shifts of this type and magnitude would also arise for solid state baths.

The fidelity plots show a quite different trend with increasing $J_x$. After a short time ${\mathcal{F}}(t)$ begins to decay more 
rapidly for larger values of $J_x$. This effect is well captured by the master equation except, again, for the smallest $J_x$ values. 
In Fig. 2 (a) we plot the canonical average of the bath coupling operator. It clearly grows in magnitude with $J_x$ and so 
the shift $\hat{S}\bar{B}$, predicted by Eq. (\ref{sra}), should also grow. This is clearly in agreement with the trend in Fig. 1(b).
This sensitivity of fidelity to $J_x$ also suggests that the shift is sensitive to the bath self-interaction strength. 

\begin{figure}
\centering
\includegraphics[scale=0.3]{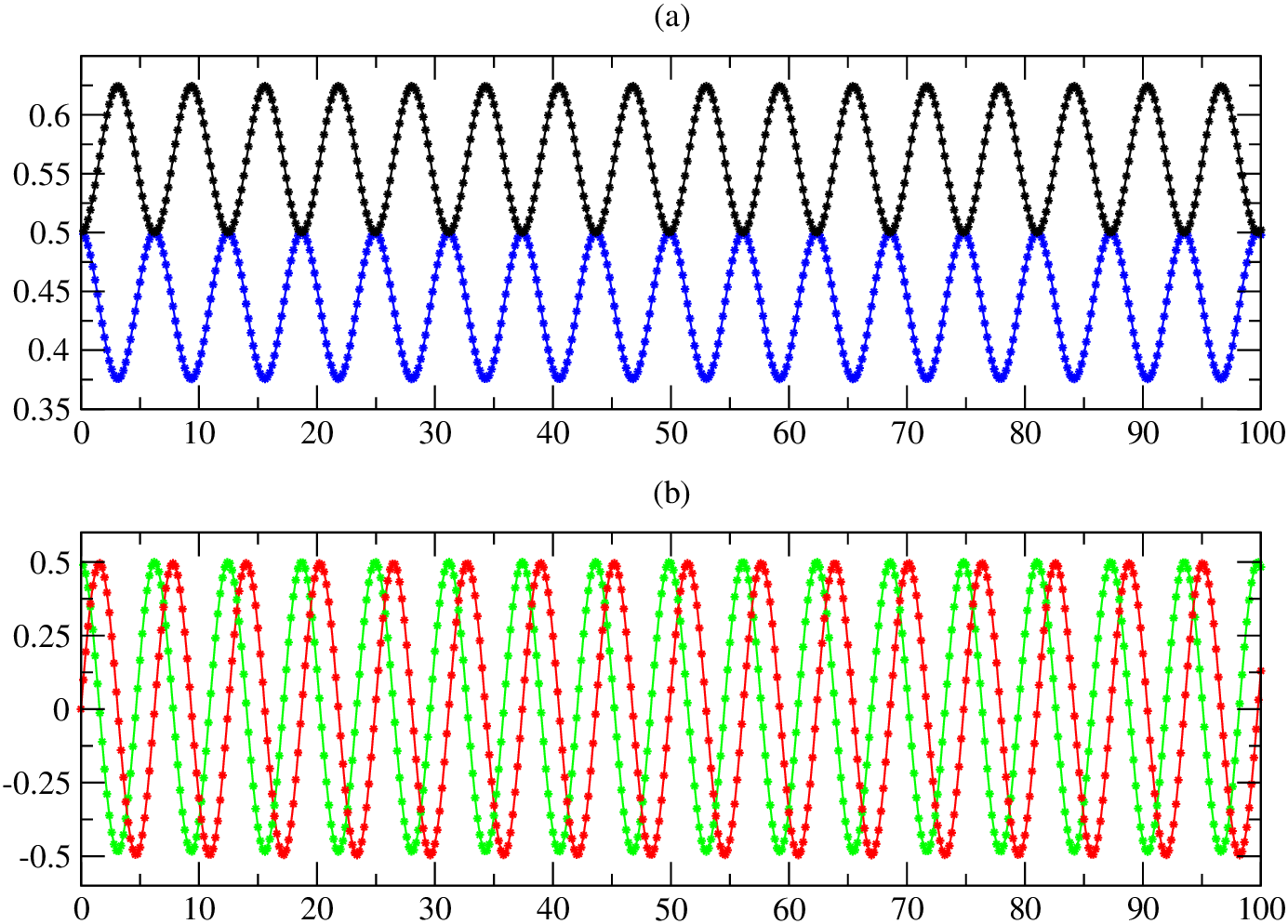} 
\caption{(Colour online) Matrix elements of system density operator $\hat{\rho}_S(t)$ for $J_{x}=1.00$: Exact numerical results are given by solid lines and master equation results by dots. (a) Diagonals of $\hat{\rho}_S(t)$, $\langle 0|\hat{\rho}_S(t)|0\rangle$ (blue) and $ \langle 1|\hat{\rho}_S(t)|1\rangle$ (black). (b) Real part of off-diagonal element of $\hat{\rho}_S(t)$, ${\rm Re}\{\langle 0 |\hat{\rho}_S(t)|1\rangle \}$ (green) and imaginary part of off-diagonal of $\hat{\rho}_S(t)$, ${\rm Im}\{\langle 0 |\hat{\rho}_S(t)|1\rangle \}$ (red).} 
\label{Fig3}
\end{figure}

In Figure 3 we plot the matrix elements of the detector qubit density $\hat{\rho}_S(t)$ for $J_{x}=1.00$. Recall that only phase evolution is expected in the absence of coherent shifting. Only very small population transfers would be expected as a result of dissipation. In Fig. 2(a) we see large Rabi oscillations in the populations which are thus a direct consequence of coherent shifting. In addition, complete agreement between the master equation and exact numerical results is evident for populations (see 3(a)) and phases (see 3(b)). This evidence strongly suggests that our choice (\ref{Ptrue}) for the projection operator $\hat{P}$ is the physically correct one.

\subsection{Very Long Time Dynamics}
In Figures 4(a) and 4(b) we plot exact numerical results for purity and fidelity, respectively, for five different values of $J_{x}$. The master equation results for the same quantities are plotted in Figures 5(a) and 5(b).

Figures 4(a) and 5(a) show that suppression of decoherence with increasing $J_{x}$ also occurs in the long time dynamics of the purity. In addition, the purity plots for the exact calculations reflect the presence of long time partial recurrences in the integrable bath regime. These recurrences are almost certainly caused by memory effects induced
because of the small bath size and
they do not appear in the master equation results. The magnitudes of the exact and ME predictions for the purity 
are however in good agreement. With increasing bath chaos the recurrences are considerably reduced and agreement
between the exact and ME results is much better.

\begin{figure}
\centering
\includegraphics*[scale=0.3]{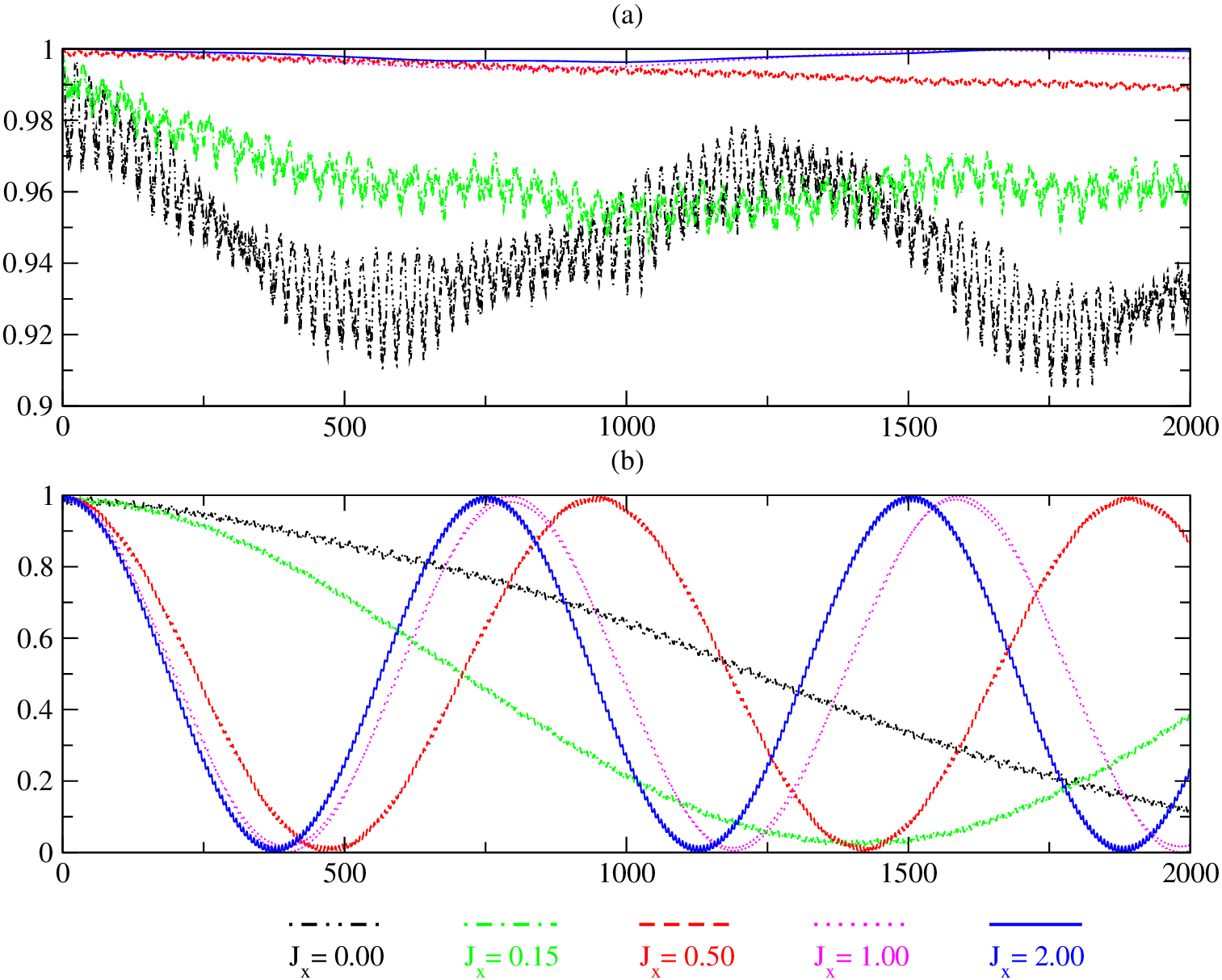} 
\caption{(Colour online) Exact numerical results for long time dynamics: (a) Purity ${\mathcal{P}}(t)$ vs. time, (b) Fidelity ${\mathcal{F}}(t)$ vs. time for $J_{x}=0.00,0.15,0.50,1.00,2.00$.}
\label{Fig4}
\end{figure}

\begin{figure}
\centering
\includegraphics*[scale=0.3]{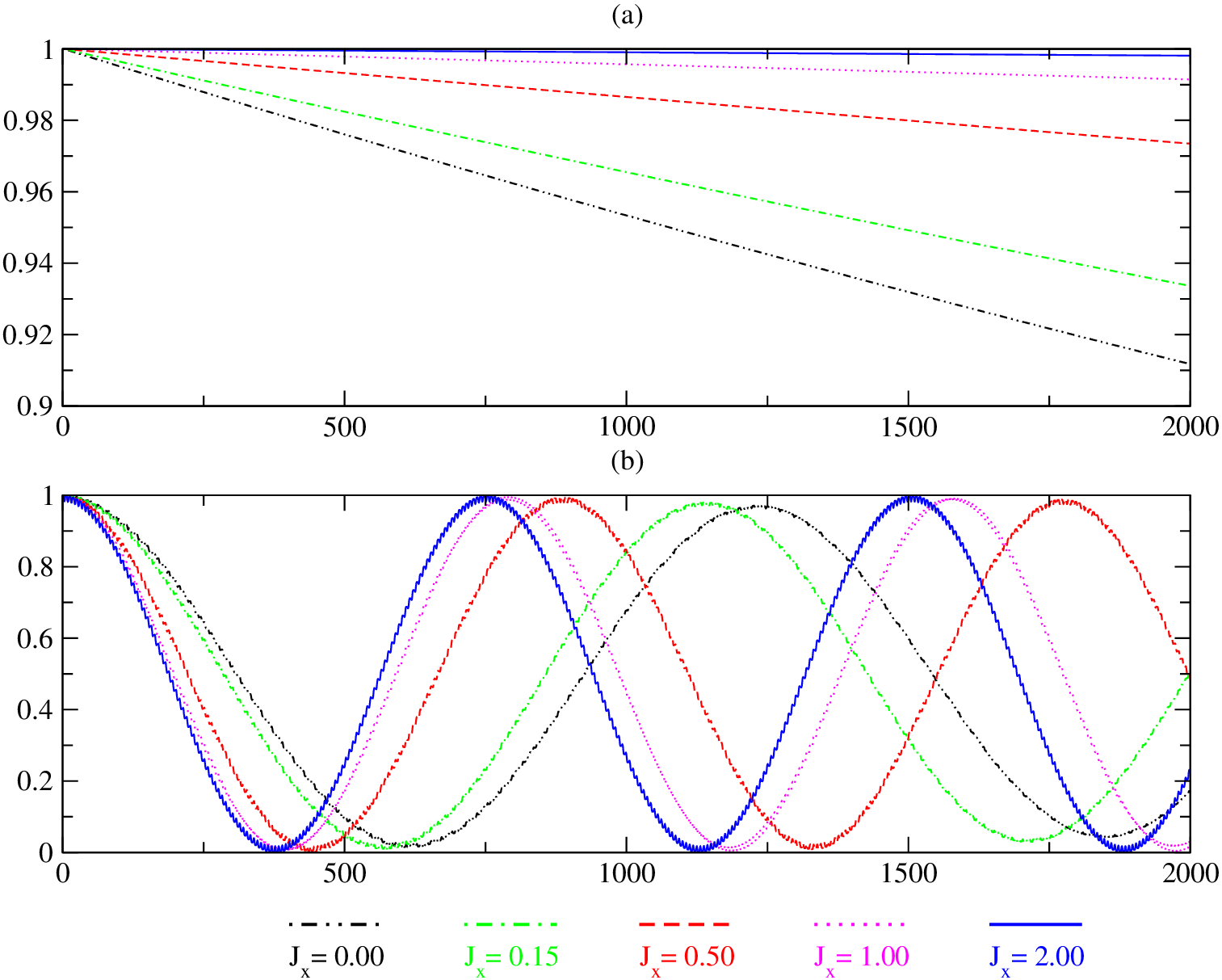} 
\caption{(Colour online) Master equation results for long time dynamics: (a) Purity ${\mathcal{P}}(t)$ vs. time, (b) Fidelity ${\mathcal{F}}(t)$ vs. time for $J_{x}=0.00,0.15,0.50,1.00,2.00$.}
\label{Fig5}
\end{figure}

In Figures 4(b) and 5(b) we plot the exact numerical results and master equation results, respectively, for the long time dynamics of the fidelity for five different values of $J_{x}$. In the long time limit, the unitary effects of system-environment interactions overwhelm the non-unitary ones, and the contributions of decoherence and dissipation to the dynamics are hardly noticeable anymore. This is true even in the integrable regime. Small magnitude, high frequency, oscillations are still noticeable. However, the fidelity plot displays an additional long time, large amplitude, periodicity. Moreover, the period of the fidelity is strongly dependent on the magnitude of $J_x$. There is again a discrepancy between 
the ME and exact results in the integrable regime, but the agreement is excellent in the chaotic cases. 

Explaining the sensitivity of the fidelity period to $J_x$ is quite straightforward.
Neglecting the effects of decoherence and dissipation it can readily be argued that the shifted system dynamics 
should beat with a frequency of $\Omega=\sqrt{\bar{B}^2+B_0^{z2}/4}/\hbar$ while the unperturbed system phase evolves with frequency $\omega=B_0^z/2\hbar$. The period of Rabi oscillations in Fig. (3) is $\pi/\Omega$. It then follows that the fidelity takes the form
\begin{eqnarray}
{\cal F}(t)&=&\frac{1}{2}[1+(\frac{\bar{B}}{\Omega})^2\cos 2\omega t-\frac{B_0^z(\Omega-B_0^z/2)}{4\Omega^2}\cos2(\Omega+\omega)t\nonumber \\
&+&\frac{B_0^z(\Omega+B_0^z/2)}{4\Omega^2}\cos2(\Omega-\omega)t].
\end{eqnarray}
The second term is very small since $\bar{B}^2$ is very much smaller than $\Omega^2$. The third term is small since $\Omega-B_0^z/2<\Omega$, and
the fourth term is of order 1. Hence, the small magnitude oscillations in fidelity have frequency $2(\Omega+\omega)$. The large amplitude oscillations are caused
by the fourth term, and they have period $\pi/(\Omega-\omega)\simeq hB_0^z/2\bar{B}^2$. Generally, $\bar{B}$ is much smaller than $B_0^z$ but $\bar{B}$ increases
with $J_x$ resulting in a shorter period. This gives rise to the changes in the period of the fidelity and fully explains the behavior observed in Figs. 4(b) and 5(b). Since $2(\Omega+\omega)$ varies more slowly with $\bar{B}$ the Rabi oscillations are not very sensitive to $J_x$. 

Thus, by measuring the period of the fidelity oscillation one can obtain an estimate of $\bar{B}$, and from Fig. 2(a) we can then obtain the 
magnitude of $J_x$. Hence, we have a detector of the strength of bath self-interaction for this JJ QC model. The same basic setup should also
carry over to the case of oscillator baths, and the technique could potentially be used to measure the strength of anharmonic interactions
in the solid state.

\section{Summary}

We studied the decoherence dynamics of a central spin interacting with a self-interacting spin bath by exact numerical calculations and a master equation approach. We obtained good agreement between the two methods. The model represents a qubit detector interacting with an isolated flawed QC. In the absence of 
interactions with the QC the detector undergoes phase evolution only. When the detector experiences a coherent shift as a result of interaction
with the QC it begins Rabi oscillations. The fidelity also exhibits a periodicity on a much longer timescale with a period which is sensitive
to the strength of bath self-interaction. Measuring the period allows one to find the intra-bath coupling strength. The same approach could be used
for measuring anharmonic interactions in more general contexts.

\ack{The authors acknowledge the support of the Natural Sciences and Engineering Research Council of Canada and  the computer resources provided by WestGrid.}

\end{document}